\documentclass[aps,prl,reprint, amsmath, amssymb, aps, groupedaddress]{revtex4-1}

\usepackage{graphicx}
\usepackage{dcolumn}
\usepackage{bm}

\begin{document}


\title{A Novel Theory for Transient Light Matter Interaction}


\author{Tomobumi Mishina}
\email{mis@phys.sci.hokudai.ac.jp}
\affiliation{Department of Physics, Faculty of Science, Hokkaido University, Sapporo 060-0810, Japan}


\date{\today}

\begin{abstract}
We propose a theory to explain the experimental results regarding short, intense optical pulses.
This theory is characterized by conjugate momentum interaction and an especially large
quantum enhancement factor and unifies the generation processes of coherent phonons, 
which was formerly attributed to impulsive stimulated Raman scattering and
displacive excitation of coherent phonons.
We apply the proposed theory to the instantaneous generation of lattice strain and derive
the quantitative relationship between incident fluence and resultant strain.  
\end{abstract}

\pacs{78.30.-j, 42.50.Ct, 42.65.Re, 78.47.-p}

\maketitle

\begin{figure}[bbb]
\centering
\includegraphics[width=4.0cm]{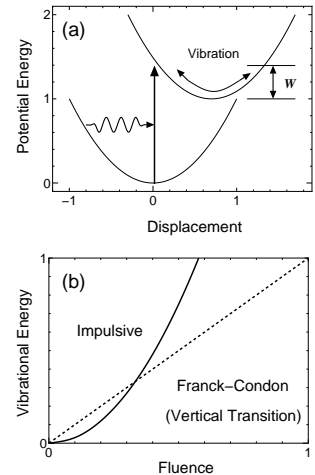}
\caption{ \label{fig:Fluence} 
(a) Schematic diagram of the DECP process based on the Franck-Condon principle. 
(b) Fluence dependence of vibrational energy under impulsive excitation.
Because the phonon amplitude is found experimentally to be proportional to the fluence,
the vibrational energy is quadratic in fluence whereas the DECP process is linear in fluence.}
\end{figure}

The Franck-Condon principle \cite{Franck26, Condon26, Birge26} is an intuitive semiclassical approach
and is widely applied various light-matter interactions.
Many spectral-domain experiments under weak continuous excitation
are well explained by the Franck-Condon principle \cite{Friedrich84}.
Advances in ultrafast laser technology enable the impulsive generation
and detection of lattice vibrations in the time domain (the so-called coherent phonons).
A large number of coherent-phonon experiments \cite{Silvestri85, Cho90, Garrett96, Merlin97, Mishina00}
have been conducted using intense transient excitation. 
To date, the generation mechanism of coherent phonons has been commonly attributed to
displacive exciteation of coherent phonons (DECP) and impulsive stimulated Raman scattering (ISRS).
Although the experimental conditions involved drastically differ, these mechanisms are still based
on the Franck-Condon principle and similar concepts.     
The DECP process is shown schematically in Fig.~\ref{fig:Fluence}(a). 
Pulsed optical excitation causes the transition between adiabatic potentials without
changing the atomic position (Franck-Condon principle), and the atom initiates the coherent-phonon oscillation 
in the electronic excited state. In this model, the mechanical energy $W$ is proportional to optical fluence.
Conversely, experimental results clearly show that the amplitude of coherent phonons is
proportional to optical fluence. Because the vibrational energy is proportional to the
square of its amplitude, the mechanical energy of coherent phonons should be proportional to the square of
optical fluence, as shown in Fig.~\ref{fig:Fluence}(b).
This discrepancy in the dependence of coherent phonons on fluence suggests that
a different approach is needed to understand transient
light-matter interactions under short, intense optical excitation.  

In this study, we introduce an effective Hamiltonian that can solve this problem and
describe the resultant light-matter interactions.
We apply this Hamiltonian to an ideal lattice composed of atomic layers and
obtain a quantitative explanation of nonthermally generated strain.

First, we consider the motion of a nucleus in a lattice system. 
Let $\hat{q}$, $\hat{p}$, $m$, and $Z$ be the position coordinate operator,  conjugate momentum operator,
mass, and atomic number of the nucleus, respectively. 
The Hamiltonian of the nucleus interacting with an electromagnetic field is 
\begin{equation}
\label{Nucleus Hamiltonian}
\hat{\cal H}_{\text{nuc}}(\hat{q},\hat{p})=\frac{1}{2m} \left( \hat{p}-ZeA(\hat q) \right) ^2+Ze\phi(\hat q),
\end{equation}
where $e$,  $A$, and $\phi$ are the elementary charge and the vector and scalar electromagnetic potentials, respectively.
The potentials include the effect of the lattice system, which is composed of electrons and the other nuclei, 
and the effect of the external electromagnetic field. When a resonant optical field is applied to the lattice system,
the electromagnetic potentials become very complex.    
Apart from this bare electromagnetic interaction, the effective Hamiltonian for transient light-matter interactions is
simply introduced by inspection. The Hamiltonian should have the position coordinate of the nucleus and its conjugate
momentum; moreover, it should reproduce the phonon amplitude proportional
to the optical fluence around the equilibrium position.      
Therefore, the form of the effective Hamiltonian is     
\begin{equation}
\label{Hamiltonian Form}
\hat{\cal H}_{\text{form}}(\hat{q},\hat{p})=\frac{1}{2m}\hat{p}^2+\frac{k}{2}{\hat q}^2
+\left( -\xi \hat{q}+\eta \hat{p} \right) I(t),
\end{equation}
where $\xi$, $\eta$, and $k$ are the optical-interaction coefficients and
the spring constant around the equilibrium position, respectively.   
The intensity $I(t)$ of light is expressed as the product of the velocity and 
the energy density inside the lattice system:
\begin{equation}
\label{Energy Flux}
I(t)=\frac{c}{n} \cdot n^2 \frac{\varepsilon_0 \overline{E^2(t)}}{2}.
\end{equation}
In Eq.~(\ref{Energy Flux}), $c$ and $\varepsilon_0$ are the speed of light and the dielectric constant in vacuum,
and $n$ is the real part of the complex index of refraction of the crystal lattice.
The square of the electric filed $E(t)$ is averaged over one optical cycle.
In addition, the time derivative of the mechanical energy $W(t)$ of the nucleus
\begin{equation}
\label{Mechanical Energy}
\frac{\text{d}}{{\text{d}}t} W(t)=\left( \frac{p(t)}{m}\xi + kq(t)\eta \right) I(t)
\end{equation}
is obtained from Eq.~(\ref{Hamiltonian Form}), where $p(t)$ and $q(t)$
are the expectation values of $\hat p$ and $\hat q$, respectively.
If Eq.~(\ref{Hamiltonian Form}) is regarded as an externally driven harmonic oscillator
with eigenfrequency $\Omega=\sqrt{k/m}$, the second quantized Hamiltonian is  
\begin{equation}
\label{Second Quantized Hamiltonian}
\hat{\cal H}_{\text{sq}}=\frac{\hbar\Omega}{2}
\left( \hat{a}^\dagger\hat{a}+\hat{a}\hat{a}^\dagger\right)
-\frac{I(t)}{2}\left( \Xi{\hat a}+\Xi^*{\hat a}^\dagger   \right).\\
\end{equation}
The annihilation operator $\hat{a}$ and the complex interaction coefficient $\Xi$ are defined as 
\begin{eqnarray}
\label{Annihilation Operator}
\hat a&&=\sqrt{\frac{m\Omega}{2\hbar}}\hat q + i\sqrt{\frac{1}{2m \hbar \Omega}}{\hat p},\\
\label{Coupling Coefficient}
\Xi&&=\sqrt{\frac{2\hbar}{m\Omega}}\xi+i\sqrt{2m\hbar\Omega}\eta.
\end{eqnarray}
By using Eq.~(\ref{Second Quantized Hamiltonian}) and the commutation relation $\left[ \hat{a},\hat{a}^{\dagger} \right]=1$,
the equation of motion for the expectation value  $a(t)$ of the annihilation operator $\hat{a}$ is
\begin{equation}
\label{Equation of Motion}
\frac{\text{d}} { {\text{d}} t }a(t)+i\Omega a(t)=-\frac{\Xi^{*}}{2i\hbar}I(t),
\end{equation}
which is easily integrated as
\begin{equation}
\label{Integration}
a(t)=-\frac{\Xi^{*}}{2i\hbar}\int_{-\infty}^{t}e^{i\Omega(t'-t)}I(t')\text{d}\it{t'}.
\end{equation}
If we define the time-dependent variables $C_{\text{R}}(t)$ and $C_{\text{I}}(t)$ by 
\begin{equation}
\label{Integration2}
C_{\text{R}}(t)+iC_{\text{I}}(t) \equiv \int_{-\infty}^{t}e^{i\Omega(t'-t)}I(t')\text{d}\it{t'},
\end{equation}
the expectation values of $\hat{q}$ and $\hat{p}$,
\begin{subequations}
\begin{eqnarray}
q(t) &&=\eta C_{\text{R}}(t) -  \frac{\xi}{m\Omega}C_{\text{I}}(t),\\
p(t) &&=\xi C_{\text{R}}(t) + m \Omega \eta C_{\text{I}}(t),
\end{eqnarray}
\end{subequations}
are obtained from Eq.~(\ref{Annihilation Operator}).
Note that, in the impulsive limit, $C_{\text{R}}(t)$ and $C_{\text{I}}(t)$ have cosine and sine waveforms, respectively.
Therefore, coherent-phonon-generation processes, which have been classified so far as DECP or ISRS mechanisms,
are unified in this theory as a transient light-matter interaction. 

\begin{figure}[bbb]
\centering
\includegraphics[width=4.0cm]{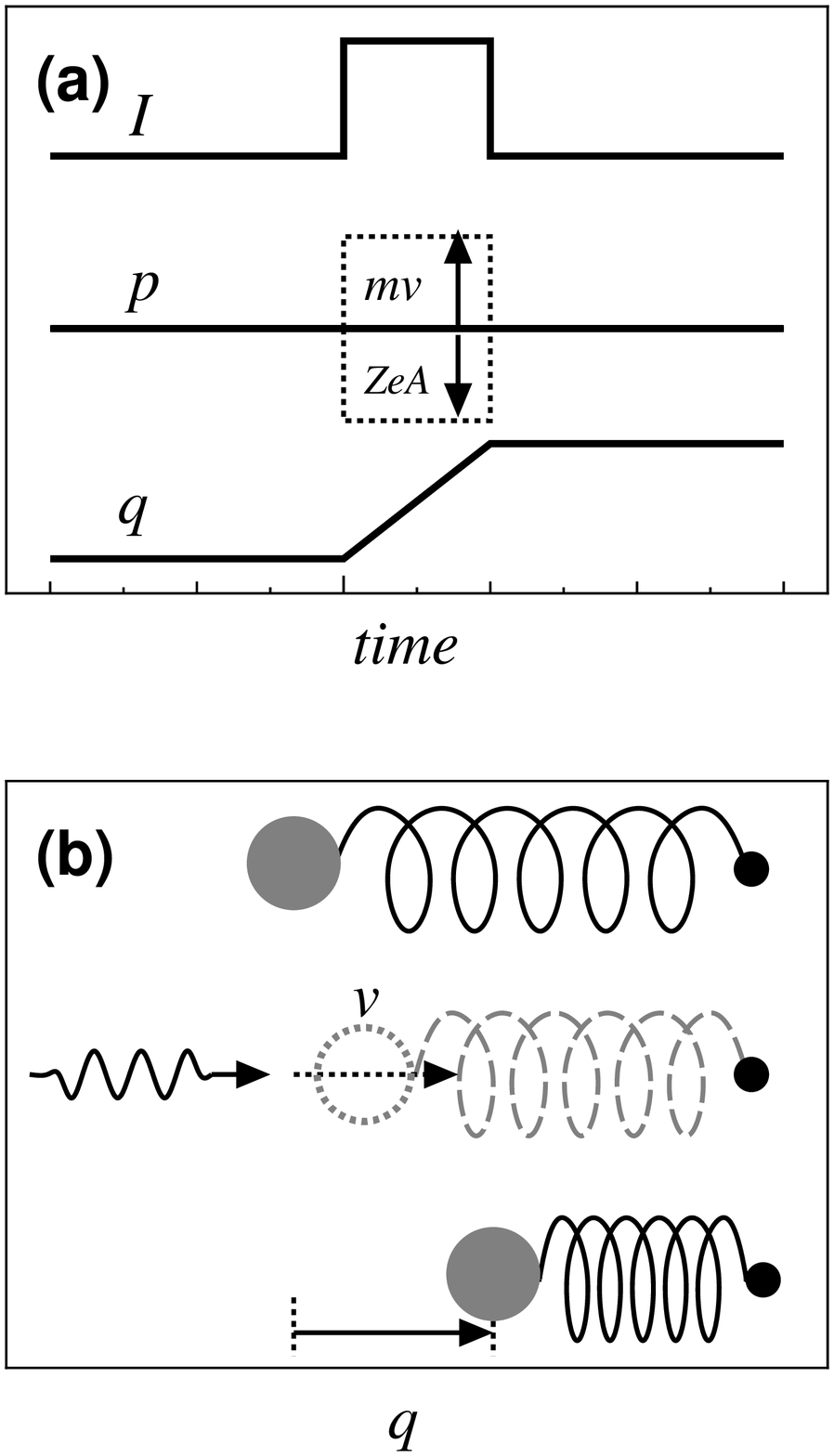}
\caption{ \label{Quantum Motion} 
(a) Time evolution of the nucleus in the impulsive limit through conjugate momentum interaction.
The total momentum remains unchanged whereas the kinetic momentum is temporally supplied by
the electromagnetic field.  
(b) The nucleus moves from one rest position to another rest position. The absorbed photon energy 
is internally converted to mechanical potential energy. }
\end{figure}

Let us evaluate the time evolution of nucleus motion in the impulsive limit.
Because the effect of the coupling constant $\xi$ has already been discussed in many articles 
\cite{Merlin97, Stevens02},
we concentrate here on the coupling coefficient $\eta$ that has been introduced in this work.
Hereafter, we call this interaction the conjugate momentum interaction.  
The time evolution of the intensity of the laser beam is   
\begin{equation}
\label{Pulse Shape}
I(t)=I_{0}\left\{ \theta(t)-\theta(t-\tau) \right\},
\end{equation}
where $\theta$ is a step function and the duration time $\tau$ is 
$(0 \leq \tau \ll 1/\Omega)$.
If we assume $\xi$ is 0, then
\begin{subequations}
\begin{eqnarray}
p(t) && \approx 0, \\
q(t) && \approx \eta I_{0} t. 
\label{Impulsive Response}
\end{eqnarray}
\end{subequations}
The time evolution is shown schematically in Fig.~\ref{Quantum Motion}(a). 
This nucleus motion is somewhat controversial because the nucleus moves whereas the momentum is always 0. 
However, this is not surprising because the conjugate momentum is a sum of the nucleus momentum and
of the momentum of the electromagnetic field. 
During the interaction, the nuclear momentum is supplied by the electromagnetic field, and it is returned to the field
when the interaction finishes.
The transient momentum $ZeA$ could be much larger than the momentum of the incident optical field.
The nucleus moves from one rest position to another rest position and
the elastic energy held in the electron system increases, as shown
in Fig.~\ref{Quantum Motion}(b).    
In the whole process, the absorbed photon energy is internally converted to elastic energy. 
Therefore, energy and momentum conservation are not violated.

The goal of establishing the effective Hamiltonian is to determine
the coupling constant by using the physical parameters of the lattice. 
If we define the electron optical polarizability around the nucleus as $\chi$, the resultant dielectric
energy $U=\chi \overline{E^{2}(t)}/2$. 
By assuming that the driving term of Eq.~(\ref{Second Quantized Hamiltonian}) is proportional
to the potential force derived from dielectric energy \cite{Merlin97},  
\begin{equation}
\label{Driving Force}
\Xi I(t) = g \frac{{\text{d}}U} {{\rm d}a}=g \frac{{\text{d}\chi}} {{\text{d}} a} \frac{\overline {E^{2}(t) } } {2},
\end{equation}
where $g$ is a nondimensional coupling constant. 
By using Eq.~(\ref{Energy Flux}), Eq.~(\ref{Driving Force}) is rewritten as
\begin{equation}
\Xi = \frac{g}{n c\varepsilon_0}\frac{\text{d}\chi}{{\text{d}} a}.  
\end{equation}
Because $\chi$ is an analytic function of complex variables, the Cauchy-Riemann  equations are
\begin{subequations}
\begin{equation}
\left( \frac{\text{d}\chi}{{\text{d}} a} \right)_{\text{Re}}=\frac{\partial \chi_{\text{Re}}}{\partial a_{\text{Re}}}
=\frac{\partial \chi_{\text{Im}}}{\partial a_{\text{Im}}},
\end{equation}
\begin{equation}
\left( \frac{\text{d}\chi}{{\text{d}} a} \right)_{\text{Im}}=\frac{\partial \chi_{\text{Im}}}{\partial a_{\text{Re}}}
=-\frac{\partial \chi_{\text{Re}}}{\partial a_{\text{Im}}}.
\end{equation}
\end{subequations}
From Eqs.~(\ref{Annihilation Operator}) and (\ref{Coupling Coefficient}), the optical-interaction coefficients are  
\begin{subequations}
\label{Interaction Coefficients}
\begin{eqnarray}
\xi && = \frac {g} {n c\varepsilon_0}\frac{\partial \chi_{\text{Re}}}{\partial q}
=\frac{gm\Omega}{n c\varepsilon_0}\frac{\partial \chi_{\text{Im}}}{\partial p}, \\
\eta && = -\frac {g}{n c\varepsilon_0}\frac{\partial \chi_{\text{Re}}}{\partial p}
=\frac{g}{m\Omega n c\varepsilon_0}\frac{\partial \chi_{\text{Im}}}{\partial q}.
\end{eqnarray}
\end{subequations}
Without specific knowledge about the bonding electron system, this simple relationship effectively links
the photoabsorption and the quantum motion of nuclei as follows.
If we insert the parameters into Eq.~(\ref{Mechanical Energy}) with the imaginary part of
Eqs.~(\ref{Interaction Coefficients}), we obtain
\begin{equation}
\label{Mechanical Energy2}
\frac{\text{d}}{{\text{d}}t}W(t)=\frac {g\Omega}{n c\varepsilon_0}
\left( \frac{\partial \chi_{\text{Im}}}{\partial q} q(t)+\frac{\partial \chi_{\text{Im}}}{\partial p} p(t)\right)I(t).
\end{equation}
In general, the rate of photon-energy absorption is expressed by the electron optical polarizability, which is
rewritten by using Eq.~(\ref{Energy Flux}) as     
\begin{equation}
\label{Photo Absorption Rate}
\frac{1}{2}\omega \chi_{\text{Im}} \overline{E^2(t)}
=\frac {\omega \chi_{\text{Im}}} {n c\varepsilon_0 }I(t),
\end{equation}
where $\omega$ is the optical frequency of the laser pulse. 
The Taylor expansion of the imaginary part of the polarizability is
\begin{equation}
\label{Imaginary Polarizability}
\chi_{\text{Im}}=\chi_{\text{Im}}^{(0)}+\frac{\partial \chi_{\text{Im}} }{\partial q}q(t)+\frac{\partial \chi_{\text{Im}}}{\partial p}p(t).
\end{equation}
The first term on the right-hand side corresponds to normal photoabsorption, and the other terms are responsible
for transient light-matter interactions. 
The resulting change in photoabsorption represents a reaction of the bonding electrons to the quantum motion of the nuclei
and guarantees that the amplitude of coherent phonons is linear in optical fluence, which cannot be explained
by semiclassical theory.
If we insert Eq.~(\ref{Imaginary Polarizability}) into Eq.~(\ref{Photo Absorption Rate})
and compare the result with Eq.~(\ref{Mechanical Energy2}), energy conservation requires
\begin{equation}
\label{Coupling Constant}
g=\omega/\Omega.
\end{equation}
Under typical experimental conditions, the phonon frequency $\Omega$ is located in THz frequency range,
whereas the laser light is in the near-visible range. Therefore, the coupling constant $g$ takes the value of several hundred,
which leads to a very efficient enhancement of transient light-matter interactions. 

From a theoretical point of view, transient light-matter interactions are considered as a
special example of a "finite-size correction of Fermi's golden rule," which has been recently proposed \cite{Ishikawa15}.
In Fermi's golden rule, which governs normal photoabsorption, energy conservation is enforced by a delta function,
such as $\delta (\omega - \Omega)$, and a constant transition probability is defined.  
However, for finite-size corrections, energy conservation is expressed by the ratio of energy quanta,
such as $\omega/\Omega$. 
By using the real part of  Eqs.~(\ref{Interaction Coefficients}), the final form of the effective Hamiltonian for transient
light-matter interactions is
\begin{equation}
\label{Effective Hamiltonian}
\hat{\cal H}_{\text{eff}}=\frac{1}{2m}\hat{p}^2+\frac{k}{2}{\hat q}^2
-\frac{\omega}{\Omega}
\left( \frac{\partial \chi_{\text{Re}}}{\partial q} \hat{q}+\frac{\partial \chi_{\text{Re}}}{\partial p} \hat{p}\right)
\frac{I(t)}{n c\varepsilon_0}.
\end{equation}

\begin{figure}[bbb]
\centering
\includegraphics[width=6.0 cm]{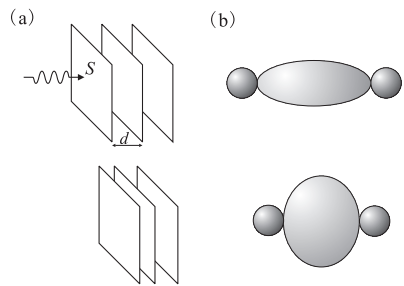}
\caption{ \label{Strain Generation} 
(a) Ideal lattice consisting of atomic layers for considering strain-pulse generation
through the conjugate momentum interaction.
(b) Change in wave function of bonding electrons caused by lattice displacement. }
\end{figure}

Finally, we apply this transient light-matter interaction to strain-pulse generation
by using an ideal lattice consisting of atomic layers, as shown
in Fig.~\ref{Strain Generation}(a). 

The guiding principle to consider the quantum motion of a multilayered system is
the balance of mechanical energy and photoabsorbed energy in each layer.
Let $\hat Q$,  $\hat P$ be 
the displacement of the interlayer distance
and the conjugate momentum operators, respectively, of an atomic layer whose cross-sectional area is $S$.
In this case, the monolayer Hamiltonian derived from Eq.~(\ref{Effective Hamiltonian}) is
\begin{eqnarray}
\label{Layer Hamiltonian}
\hat{\cal H}_{\text{L}}=\frac{\hat{P}^2}{2\rho Sd}
&&+\frac{YSd}{2} \left( \frac{\hat Q}{d}\right) ^2\nonumber \\
&&-\frac{\omega Sd}{\Omega_{\text{L}}}
\left( \frac{\partial \tilde{\varepsilon}_{\text{Re}}}{\partial Q} \hat{Q}
+\frac{\partial \tilde{\varepsilon}_{\text{Re}}}{\partial P} \hat{P}\right)
\frac{I(t)}{ n c},
\end{eqnarray}
where, $\rho$, $d$, $Y$, and $\tilde{\varepsilon}$ are the density, lattice constant, elastic constant, and
relative dielectric constant, respectively.
The corresponding phonon energy $ \Omega_{\rm{L}}=\sqrt{Y/\rho}/d$.
The relative dielectric constant is
\begin{equation}
\label{dielectric constant}
\tilde{\varepsilon}=\frac{1}{\varepsilon_0} \frac{\chi_{\text{L}}}{Sd},
\end{equation}
where $\chi_\text{L}$ is the total electron optical polarizability per layer.      
Under impulsive excitation, the conjugate momentum interaction predicts instantaneous generation
of lattice strain $u=Q/d$, as indicated by Eq.~(\ref{Impulsive Response}).
By using the lattice parameters and optical fluence $F(\approx I_{0}\tau)$,
the associated stress $\sigma$ is       
\begin{equation}
\label{Lattice Stress}
\sigma=Yu=\frac{ 1 } {n} \frac{\partial \tilde{\varepsilon}_{\text{Im}} }{\partial u}
\frac{2\pi F}{\lambda},
\end{equation}
where $\lambda$ is the wavelength of the laser pulse in vacuum. For simplicity, 
the real part of the index of refraction $n$ is assumed to be constant.  
Although Eq.~(\ref{Lattice Stress}) indicates the instantaneous generation
of strain without thermal expansion or high-density carriers,  
it is very difficult to verify this effect experimentally.
Low-density excitation experiments \cite{Baumberg97, Kasami04, Matsuda05} prove only
that the strain pulse is proportional to fluence.
Time-resolved diffraction experiments with x-rays and electrons \cite{Reis01, Carbone08, Raman08} under high-density
excitation reveal surprisingly large lattice displacements; however,
insufficient temporal and spatial resolution obscures the effect.
Conversely, the strain modulation of coherent phonons in antimony \cite{Kumagai16} could provide
additional information about the effect.
The reduction of the frequency of the coherent phonon immediately after the high-density pulse excitation
can be explained by the lattice strain. The experimental result shows that the
fluence of 1.0 mJ/cm$^{2}$ at a wavelength of 800 nm generates a pressure of 0.9 GPa.
The corresponding optical parameters of antimony,  $n$ and $\tilde{\varepsilon}_\text{Im}$,
are 2.50 and 23.9, respectively \cite{LandoltOpt}. 
If we evaluate the partial differential coefficient
in Eq.~(\ref{Lattice Stress}) by inserting these parameters,    
the ratio $\frac{\partial \tilde{\varepsilon}_\text{Im}}{\partial u}/\tilde{\varepsilon}_\text{Im}$
is estimated to be 1.20. 
This value indicates that the lattice strain and the relative change in the imaginary part of the
dielectric constant are of the same order of magnitude.
This result is quite reasonable because the wave function of bonding electrons depends on lattice strain,
as shown in Fig.~\ref{Strain Generation}(b).

In conclusion, we introduce herein a new effective Hamiltonian with which we study
the transient light-matter interaction induced by short, intense optical pulses.
The Hamiltonian includes the conjugate momentum interaction and
a very large quantum enhancement factor that reflects
the "finite size correction of Fermi's golden rule." 
The theory unifies the generation processes of coherent phonons formerly attributed to the  
ISRS and DECP mechanisms. 
Moreover, the theory quantitatively predicts the nonthermal generation of lattice strain
and is consistent with the experimental result in antimony.
This simple theory overturns the semiclassical concept applied heretofore in most experiments
with short, intense optical pulses.

The author thanks Professor Kenzo Ishikawa for his encouragement and stimulating discussion.

\bibliography{TLMI}

\end{document}